\documentclass[a4paper,11pt]{article}

\usepackage{pos}
\usepackage{caption}
\usepackage{subcaption}
\usepackage{lineno}
\usepackage{hyperref}

\title{High-rate electron detectors to study Compton scattering in non-perturbative QED}
\ShortTitle{High-rate electron detectors}

\author*[a,b]{Antonios Athanassiadis}
\author[a,c]{John Hallford}
\author[a]{Louis Helary}
\author[a,c]{Luke Hendriks}
\author[a]{Ruth Magdalena Jacobs}
\author[a]{Jenny List}
\author[b]{Gudrid Moortgat-Pick}
\author[a]{Evan Ranken}
\author[a]{Stefan Schmitt}
\author[a,c]{Matthew Wing}

\affiliation[a]{Deutsches Elektronen-Synchrotron DESY,\\
  Notkestr. 85, 22607 Hamburg, Germany}
\affiliation[b]{Universität Hamburg,\\
Luruper Chaussee 149, Hamburg, Germany}
\affiliation[c]{University College London (UCL),\\
Gower Street, London, UK}

\emailAdd{antonios.athanassiadis@desy.de}

\abstract{
Research in non-perturbative QED in strong-field backgrounds has gained
interest in recent years, due to advances in high-intensity laser
technologies that make extreme fields accessible in the laboratory. One
key signature of strong-field QED is non-linear Compton scattering in
collisions between a relativistic electron beam and a high-intensity laser
pulse. In the vicinity of strong fields, the electron gains a larger
effective mass, which leads to a laser-intensity-dependent shift of the
kinematic Compton edge and the appearance of higher-order harmonics in the
energy spectrum. One of the challenges of measuring the Compton energy
spectrum in laser-electron-beam collisions is the enormous flux of
outgoing Compton-scattered electrons and photons, ranging from $10^3$ to $10^9$
particles per collision. We present a combined detector system for
high-rate Compton electron detection in the context of the planned LUXE
experiment, consisting of a spatially segmented gas-filled Cherenkov
detector and a scintillator screen imaged by an optical camera system. The
detectors are placed in a forward dipole spectrometer to resolve the
electron energy spectrum. Finally, we discuss techniques to reconstruct
the non-linear Compton electron energy spectrum from the high-rate
electron detection system and to extract the features of non-perturbative
QED from the spectrum.
}

\FullConference{The European Physical Society Conference on High Energy Physics (EPS-HEP2023)\\
 21-25 August 2023\\
Hamburg, Germany\\}

\newcommand{\ValueWithUnit}[2]{#1\,\mathrm{#2}}
\newcommand{\ValueWithIndex}[2]{#1_{\mathrm{#2}}}

\newcommand{\degree}{^{\circ}}
\begin{document}
\maketitle

\section{The LUXE Experiment}
\noindent The \textbf{L}aser \textbf{U}nd \textbf{X}FEL \textbf{E}xperiment (LUXE) is a new experiment planned at DESY Hamburg, Germany \cite{CDR}. Its goal is to measure non-linear phenomena in the strong-field regime of Quantum Electrodynamics (QED) with high precision, thus allows to accurately probe theoretical predictions of the transition between perturbative and non-perturbative QED. \\

In order to create strong fields necessary for non-linear processes, highly relativistic electrons from the European XFEL will interact with a high intensity laser ($\ValueWithUnit{350}{TW}$). The laser will induce an electric field, which, due to the large Lorentz boost of the $\ValueWithUnit{16.5}{GeV}$ electron beam, will be larger than the so-called \textit{Schwinger limit} $(\ValueWithUnit{1.32\cdot 10^{18}}{Vm^{-1}})$. As a result of that, two major processes will \par be induced: \\
\begin{minipage}[t]{0.5\textwidth}
    \vspace{0.5em}
    \begin{itemize}
        \item Non-linear Compton scattering \\
        $\ValueWithIndex{e^{-}}{XFEL} + \ValueWithIndex{\gamma}{Laser} \xrightarrow{} \ValueWithIndex{e^{-}}{Compton} + \ValueWithIndex{\gamma}{Compton}$
    \end{itemize}
    \vspace{0.5em}
\end{minipage}
\begin{minipage}[t]{0.5\textwidth}
    \vspace{0.5em}
    \begin{itemize}
        \item Breit-Wheeler pair production \\
        $\ValueWithIndex{\gamma}{Compton} + \ValueWithIndex{\gamma}{Laser} \xrightarrow{} e^{-} + e^{+}$
    \end{itemize}
    \vspace{0.5em}
\end{minipage}

\begin{figure}[h!]
    \centering
    \begin{subfigure}[t]{0.48\linewidth}
        \centering
        \includegraphics[width=0.96\linewidth]{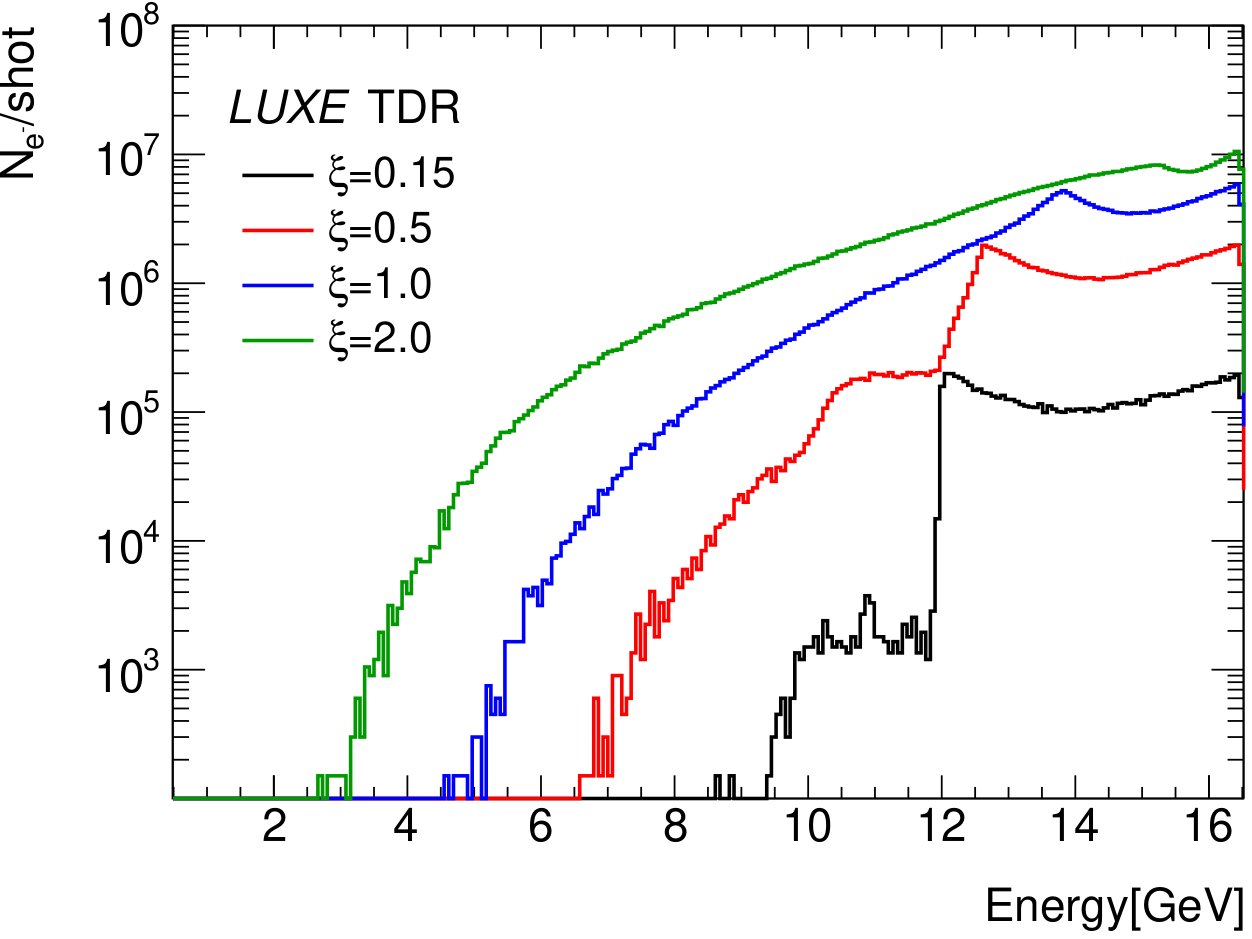}
        \caption{ }
        \label{fig:SimulationOfElectronEnergies}
    \end{subfigure}
    \hspace{1em}
    \begin{subfigure}[t]{0.48\linewidth}
        \centering
        \includegraphics[width=0.96\linewidth]{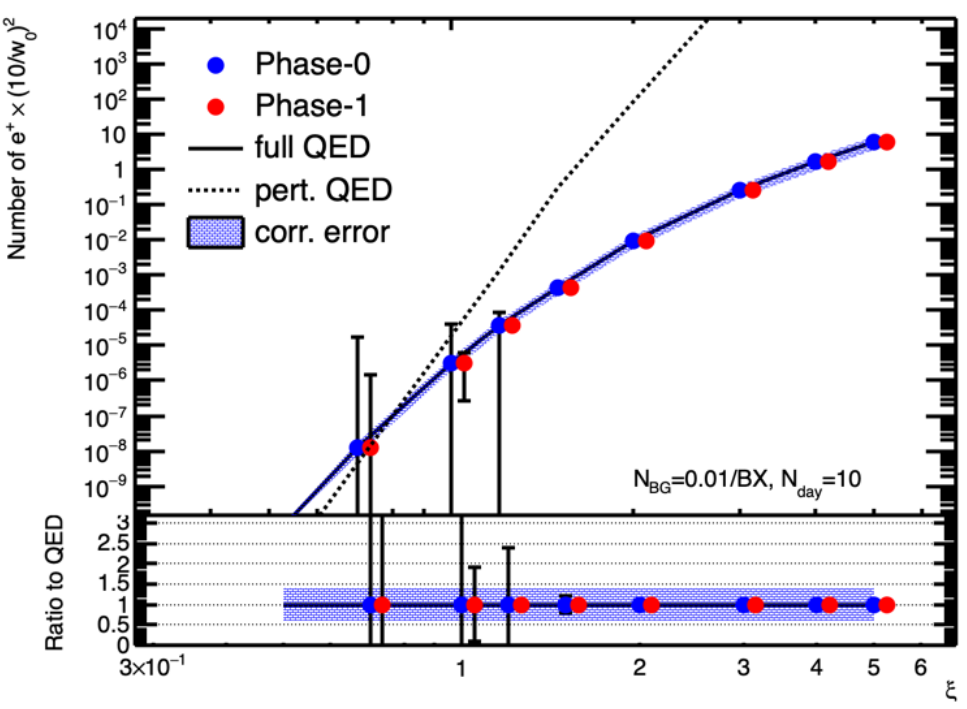}
        \caption{ }
        \label{fig:SimulationOfPositronRates}
    \end{subfigure}
    \caption{Simulation of (a) the Compton electron energy distribution and (b) the Breit-Wheeler positron production rate per bunch crossing in relation to different values of $\xi$ at LUXE \cite{CDR}.}
    \label{fig:SimulationOfPositronRatesAndElectronEnergies}
\end{figure}

The energy distribution of the Compton electrons, as well as the production rate of Breit-Wheeler pairs, depends on the non-linearity parameter $\xi$ which is defined in eq. \eqref{DefinitionOfXi}
\begin{align}
    \xi &= \frac{\ValueWithIndex{m}{e}}{\ValueWithIndex{\omega}{Laser}} \sqrt{\frac{I}{\ValueWithIndex{I}{critical}}}
    \label{DefinitionOfXi}
\end{align} 

\noindent with the electron mass $\ValueWithIndex{m}{e}$, the laser frequency $\ValueWithIndex{\omega}{Laser}$, the laser intensity $I$, and $\ValueWithIndex{I}{critical}$ the Schwinger limit. With increasing laser intensity, therefore with increasing values of $\xi$, the non-linearity increases which leads to a shift of the Compton electron energies and higher-order harmonics (Figure \ref{fig:SimulationOfElectronEnergies}) as well as an increasing number of created positrons (Figure \ref{fig:SimulationOfPositronRates}). These plots were created with \textit{Ptarmigan} \cite{ptarmigan} for the LUXE project \cite{CDR}. Remarkable are the high rates and especially the large dynamic range of particles, which will have to be measured with great accuracy.
\\

\section{LUXE Electron Detection System}
\begin{figure}[h!]
    \centering
    \includegraphics[width=0.85\linewidth]{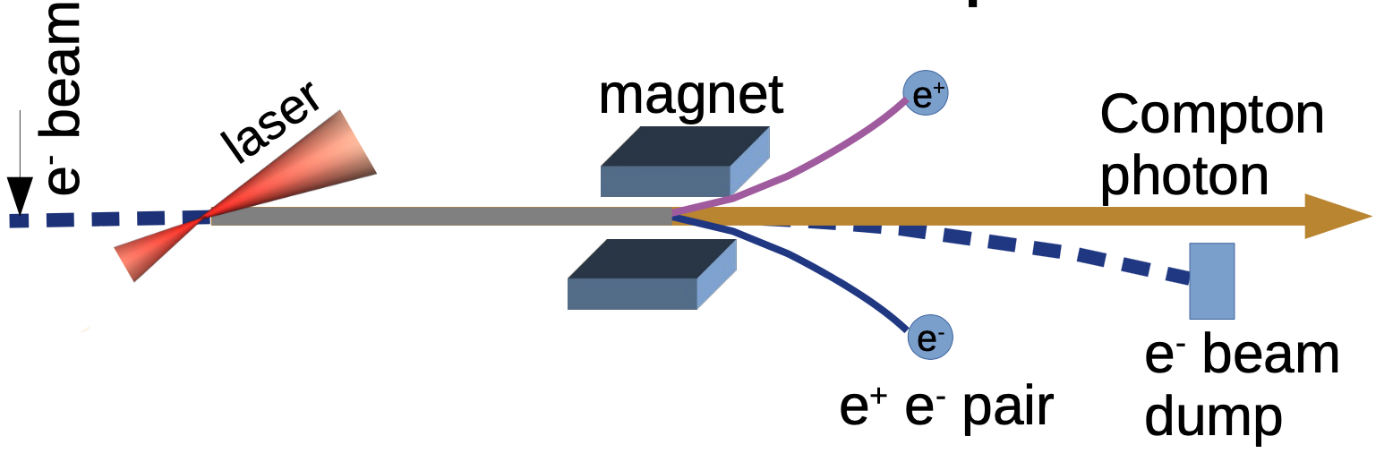}
    \caption{Schematics of the planned LUXE setup at and around the interaction point. \cite{CDR}}
    \label{fig:LUXESetup}
\end{figure}

\noindent The aim of the LUXE Electron Detection System (EDS) is to detect the electrons and measure their energy distribution with a relative resolution of under $\ValueWithUnit{2}{\%}$. The main challenge is to cover the high dynamic range of $10^3$ to $10^8$ electrons per bunch crossing and detector channel, while maintaining the linear response of the detector. As shown in figure \ref{fig:LUXESetup}, a dipole magnet is located behind the interaction point of the electron beam and the laser, which bends the trajectory of charged particles produced in the electron-laser interaction, depending on their energy.

\begin{figure}[b!]
    \centering
    \includegraphics[width=0.75\linewidth]{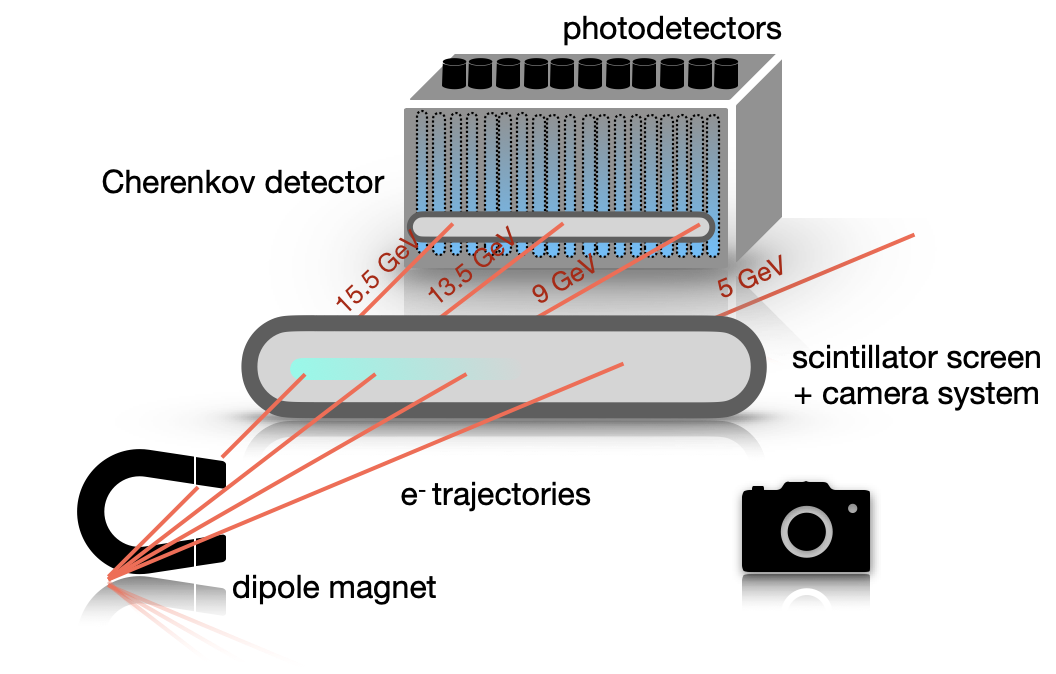}
    \caption{Schematics of the Electron Detection System placed after the dipole magnet. \cite{TDR}}
    \label{fig:EDS}
\end{figure}

In LUXE, two complementary measurements are foreseen for each detector system. For the EDS, this will be a scintillating screen and camera system in combination with a Cherenkov counter. Due to the dipoles' strong magnetic field of $\ValueWithUnit{1}{T}$, the electrons will be fanned out with respect to their energy over a distance of $\approx\ValueWithUnit{0.5}{m}$ ($\ValueWithUnit{4.3}{m}$ behind the dipole). This separation is used to measure the energy distribution with a spatially segmented detector. As shown in figure \ref{fig:EDS}, the fanned-out electrons are first detected via the scintillation screen before crossing the Cherenkov counter channels.

\subsection{The scintillator screen system}
\begin{figure}[h!]
    \centering
    \begin{subfigure}[t]{0.45\textwidth}
        \centering
        \includegraphics[width=1.25\linewidth]{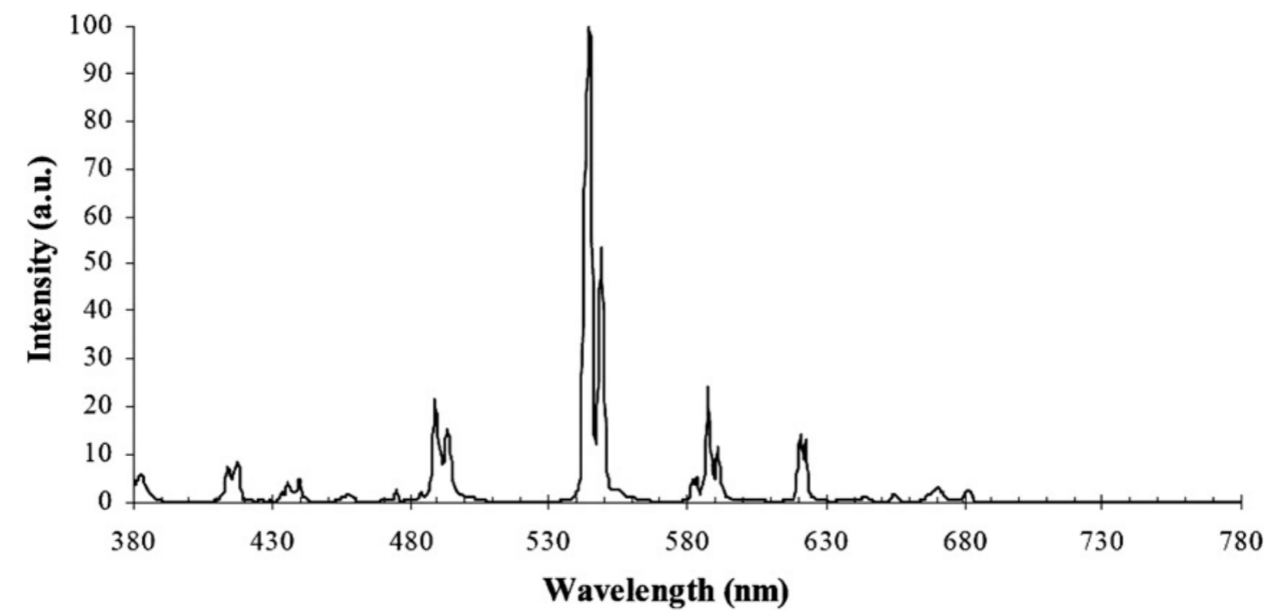}
        \caption{ }
        \label{fig:ScintiSpectrum}
    \end{subfigure}
    \hspace{1em}
    \begin{subfigure}[t]{0.45\textwidth}
        \centering
        \includegraphics[width=0.8\linewidth]{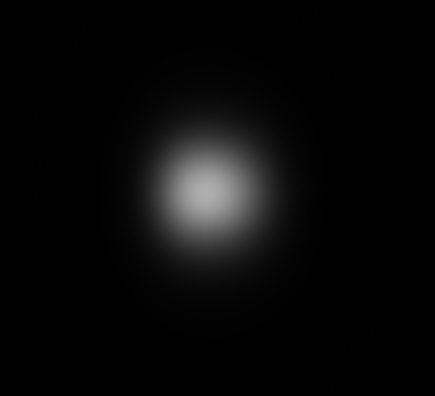}
        \caption{ }
        \label{fig:100pCScreen}
    \end{subfigure}
    \caption{(a) Wavelength spectrum of the scintillator screen. \cite{Screen} (b) Beam test: Image of $\ValueWithUnit{100}{pC}$ electrons.}
    \label{fig:ScreenProperties}
\end{figure}

\noindent The scintillating screen is made of Terbium-doped Gadolinium Oxysulfide (GadOx) \cite{Screen}. The material is chosen for the linear relation between the number of produced scintillation photons with respect to the number of incident particles and its radiation hardness of up to $\ValueWithUnit{100}{MGy}$ \cite{ScreenRadiation}. The scintillation spectrum shown in figure \ref{fig:ScintiSpectrum} indicates that a high number of scintillation photons are emitted at a wavelength of $\ValueWithUnit{543}{nm}$. Since LUXE will take data at a rate of $\ValueWithUnit{10}{Hz}$, the decay time of $\ValueWithUnit{600}{\mu s}$ is more than sufficient. The created scintillation light will be recorded with two cameras with either $2.3$ or $8.8$ Mpixel \cite{CameraQE}. In combination with optical filters for the $\ValueWithUnit{543}{nm}$, the monochromatic CMOS sensors will achieve a quantum efficiency of up to $\ValueWithUnit{70}{\%}$.
\vspace{0.25em}

The light emitted at the screen's surface will be captured by the camera system, and its intensity will be stored as a CMOS count map (or gray value per pixel). Figure \ref{fig:100pCScreen} illustrates such an image recorded with a electron bunch charge of $\ValueWithUnit{100}{pC}$ coming from the ARES facility \cite{ares}. An intensity spectrum along the horizontal axis will indicate the location of the maximum intensity. In the later experiment, this maximum position will represent the position of the Compton edge. 

\subsection{The Cherenkov detector}
\noindent The Cherenkov detector uses spatially segmented gas volumes to measure the flux of incident electrons per volume using the Cherenkov effect. Cherenkov photons un the UV and visible range are created when a charged particle moves faster than the speed of light in a medium. A material-specific threshold must be exceeded by a particle in order to create Cherenkov photons. The number of created optical photons per distance depends on the refractive index of the medium (eq. \eqref{eq:franktamm} \cite{FrankTamm})
\begin{align}
    \frac{d^2N_\gamma}{dz\,d\lambda} &= \frac{2\pi\alpha}{\lambda^2}\sin^2\theta_c \label{eq:franktamm}\\
    \mathrm{with}\quad\sin^2\ValueWithIndex{\theta}{c} &= 1 - \frac{1}{\beta^2\,n^2}.
\end{align}

\begin{figure}[h!]
    \centering
    \begin{subfigure}[t]{0.45\textwidth}
        \centering
        \includegraphics[width=0.85\linewidth]{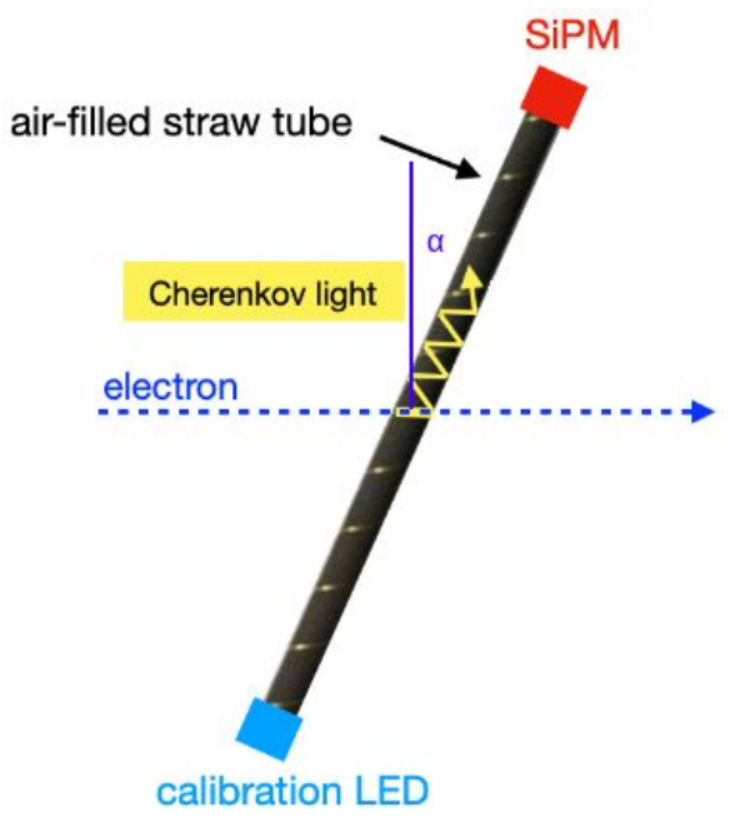}
        \caption{ }
        \label{fig:SchemeStraw}
    \end{subfigure}
    \hspace{1em}
    \begin{subfigure}[t]{0.45\textwidth}
        \centering
        \includegraphics[width=0.75\linewidth]{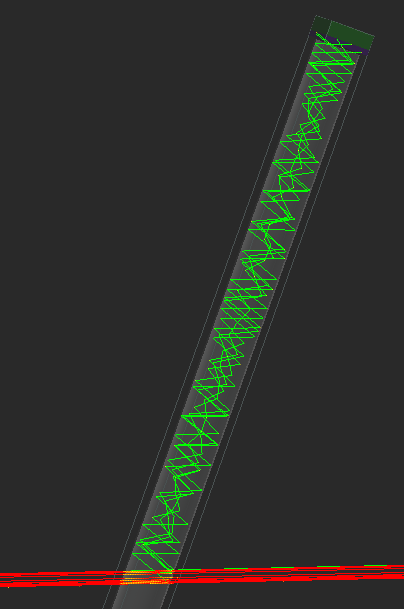}
        \caption{ }
        \label{fig:StrawGeant}
    \end{subfigure}
    \caption{(a) Schematics of a straw from the Cherenkov detector. \cite{TDR} (b) Simulation of one stainless-steel straw with $\ValueWithUnit{3}{mm}$ inner diameter and an angle of $\ValueWithUnit{20}{\degree}$ with one passing $\ValueWithUnit{150}{MeV}$ electron (red) and the track of the created optical photons (green) towards the SiPM (blue).}
    \label{fig:GEANT4Schemes}
\end{figure}

\noindent The Cherenkov detector will consist of about $200$ air-filled stainless-steel \textit{straws} (or \textit{tubes}) which will be aligned in a grid. Each straw will be equipped with a silicon photomultiplier (SiPM) as a photodetector at the top and an LED for calibration purposes at the bottom (Figure \ref{fig:SchemeStraw}). When an electron passes through the straw, the created Cherenkov light will be reflected on the inside walls of the straw towards the SiPM and then be detected. In air, with an refractive index of $1.00026$, the production threshold of about $\ValueWithUnit{22}{MeV}$ is far below the expected GeV-electron energies in LUXE but can help to reject low-energy backgrounds. 

Since the optical photons are reflected and partially absorbed by the straw walls the light yield per electron can be controlled accurately by tilting the straw with respect to the beam axis. Since the absorption probability increases with the number of reflections, a steeper tilting angle reduces the number of detectable photons.
In order to be able to characterize the planned detector system (such as SiPM characteristics as well as straw dimensions, material, reflectivity and angle), a \textit{GEANT4}\,\,\cite{geant4} simulation is currently under development. This will allow electron interactions to be studied, including scattering processes, and the optical phenomena such as the Cherenkov effect (Figure \ref{fig:StrawGeant}). This simulation also considers parameters like the straw reflectivity and angle, whose influence on the signal intensity is crucial to understand. The overall goal of optimizing the straw geometry and material properties is to match the large dynamic range that needs to be covered in order to measure the Compton spectrum in a wide range of laser intensities. \\

\newpage
\section{Future plans and conclusion}
\begin{figure}[h!]
    \centering
    \includegraphics[width=0.5\linewidth]{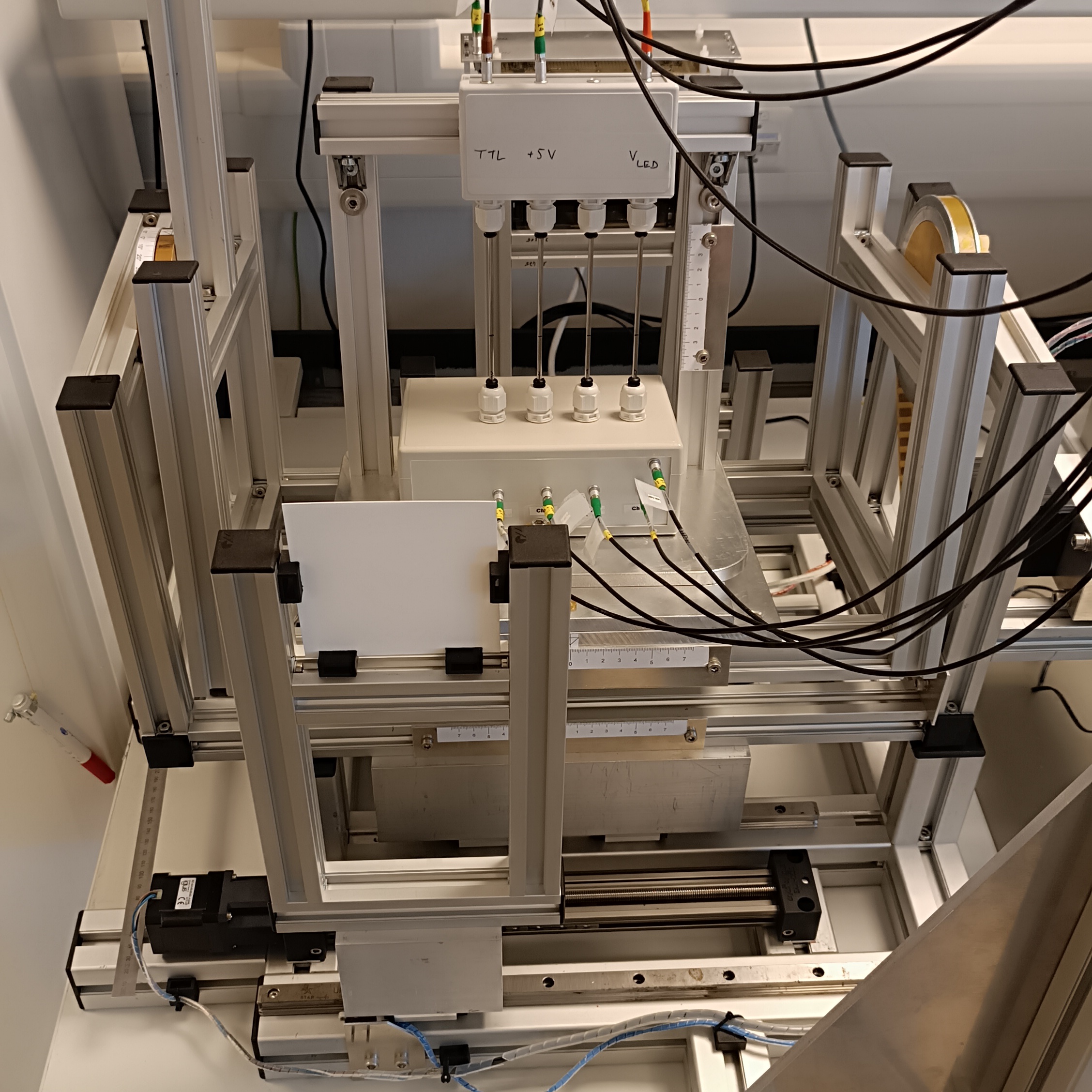}
    \caption{Image of the developed prototype setup. Electron beam direction is into the image.}
    \label{fig:Prototype}
\end{figure}

\noindent Finally, verifying simulation data, as well as testing the detector response in a real environment is a crucial part of the detector R\&D. A prototype system (Figure \ref{fig:Prototype}) is currently under test at electron accelerator facilities, such as ARES \cite{ares} at DESY. The prototype consists of a screen in front of four straws, which are read out via a camera (not seen in the image). The lower box contains various SiPM models \cite{DatasheetHamamatsu, DatasheetOnsemi}. In the upper box, UV-LEDs face inside the straw for sensor calibration. \\

LUXE demands many requirements on the detectors in order to measure the effects of strong-field QED. The large dynamic range in the expected high-flux electron regime is a challenging environment for high-resolution measurements. The two introduced detector systems are specifically developed to suit the requirements. In the future an evaluation of these beam tests, extended simulation studies as well as further development in the prototype and data acquisition will follow. \\

\small
\bibliographystyle{unsrt}

\end{document}